\documentclass[twocolumn,superscriptaddress,showpacs]{revtex4}
\usepackage{graphicx,amsmath}
\usepackage{epstopdf}
\usepackage{subfigure}

\renewcommand{\a}{$\alpha$}
\renewcommand{\b}{$\beta$}
\newcommand{\m}{$\mu$}
\newcommand{\n}{$\nu$}

\renewcommand\pm{p^{\mu}}

\newcommand\qmn{q^{\mu\nu}}

\newcommand\kn{k_{\nu}}
\newcommand\kl{k_{l}}

\newcommand\gm{G_{\mu}}

\newcommand\gmn{G_{\mu\nu}}

\newcommand\gnm{G_{\nu\mu}}

\newcommand\gmdn{G_{\mu}^{'\nu}}
\newcommand\gmdl{G_{\mu}^{'\lambda}}

\newcommand\dmn{\overline{k}^{\nu}_{\mu}}

\newcommand\uaa{u_{\alpha\alpha}}
\newcommand\ubb{u_{\beta\beta}}
\newcommand\uab{u_{\alpha\beta}}
\newcommand\uba{u_{\beta\alpha}}

\newcommand\unm{u_{\nu\mu}}
\newcommand\uom{u_{1\mu}}
\newcommand\ulm{u_{l\mu}}

\newcommand\uon{u_{1\nu}}
\newcommand\uln{u_{l\nu}}

\newcommand\xa{x_{\alpha}}
\newcommand\xb{x_{\beta}}
\newcommand\ka{k_{\alpha}}
\newcommand\kb{k_{\beta}}

\newcommand\ssa{s_{\alpha}}
\newcommand\ssb{s_{\beta}}

\newcommand\Sa{S_{\alpha}}
\newcommand\Sb{S_{\beta}}

\newcommand\Sm{S_{\mu}}

\newcommand\daa{\overline{k}^{\alpha}_{\alpha}}
\newcommand\dba{\overline{k}^{\alpha}_{\beta}}

\newcommand\dbb{\overline{k}^{\beta}_{\beta}}
\newcommand\dab{\overline{k}^{\beta}_{\alpha}}

\newcommand\dml{\overline{k}^{\lambda}_{\mu}}

\newcommand\dinter{\overline{k}_{\rm inter}}
\newcommand\dintra{\overline{k}_{\rm intra}}

\newcommand\ko{k_{1}}

\newcommand\ga{G_{\alpha}}

\newcommand\gb{G_{\beta}}

\newcommand\hm{H_{\mu}}

\newcommand\hmn{H_{\mu\nu}}

\renewcommand\hom{H_{1\mu}}

\newcommand\hlm{H_{l\mu}}

\newcommand\hlmdn{H_{\lambda\mu}^{'\nu}}
\newcommand\hgldn{H_{\gamma\lambda}^{'\nu}}

\newcommand\vecx{{\bf x}}
\newcommand\xo{x_{1}}

\newcommand\xm{x_{\mu}}
\newcommand\xn{x_{\nu}}
\newcommand\xl{x_{l}}

\newcommand\vv{x}

\newcommand\vecv{{\bf x}}
\newcommand\vm{\vv_{\mu}}
\newcommand\vn{\vv_{\nu}}

\newcommand\partvn{\frac{\partial}{\partial \vn}}

\newcommand\partxn{\frac{\partial}{\partial \xn}}

\newcommand\sm{s_{\mu}}

\newcommand\veco{{\bf 1}}

\begin{document}
\title{Percolation on interacting networks}
\author{E.~A. Leicht}
\affiliation{Department of Mechanical and Aeronautical Engineering, University of California, Davis, CA 95616}
\author{Raissa M. D'Souza}
\affiliation{Department of Mechanical and Aeronautical Engineering, University of California, Davis, CA 95616}
\affiliation{The Santa Fe Institute, Santa Fe, NM 87501}
\date{\today}

\begin{abstract}
Most networks of interest do not live in isolation. Instead they form components of larger systems in which multiple networks with distinct topologies coexist and where elements distributed amongst different networks may interact directly. 
Here we develop a mathematical framework based on generating functions for analyzing a system of $l$ interacting networks given the connectivity within and between networks.  We derive exact expressions for the percolation threshold describing the onset of large-scale connectivity in the system of networks and each network individually. These general expressions apply to networks with arbitrary degree distributions and we explicitly evaluate them for $l=2$ interacting networks 
with a few choices of degree distributions. 
We show that the percolation threshold in an individual network can be significantly lowered once ``hidden" connections to other networks are considered.  We show applications of the framework to two real-world systems involving communications networks and socio-tecnical congruence in software systems. 
\end{abstract}
%
\maketitle

In the past decade there has been a significant advance in
understanding the structure and function of networks.  Mathematical
models of networks are now widely used to describe a broad range of
complex systems, from spread of disease on networks of human contacts
to interactions amongst proteins~\cite{DM2002,mejn-03review,Boccaletti06}. 
However, current methods deal almost exclusively with
individual networks treated as isolated systems. In reality an individual network is often just one component in a much larger complex system; a system that can bring together multiple networks with distinct topologies and functions.
For instance, a pathogen spreads on a network of
human contacts abetted by global and regional transportation networks.
Likewise, email and e-commerce networks rely on the Internet which in
turn relies on the electric grid.  In biological systems, 
activated genes give rise to proteins some of which go back to the
genetic level and activate or inhibit other genes. 
Results obtained in
the context of a single isolated network can change dramatically once
interactions with other networks are incorporated.  

\begin{figure}[b]
  \begin{center}
    \subfigure[]{\label{fig:2layers}
      \includegraphics[width=.21\textwidth]{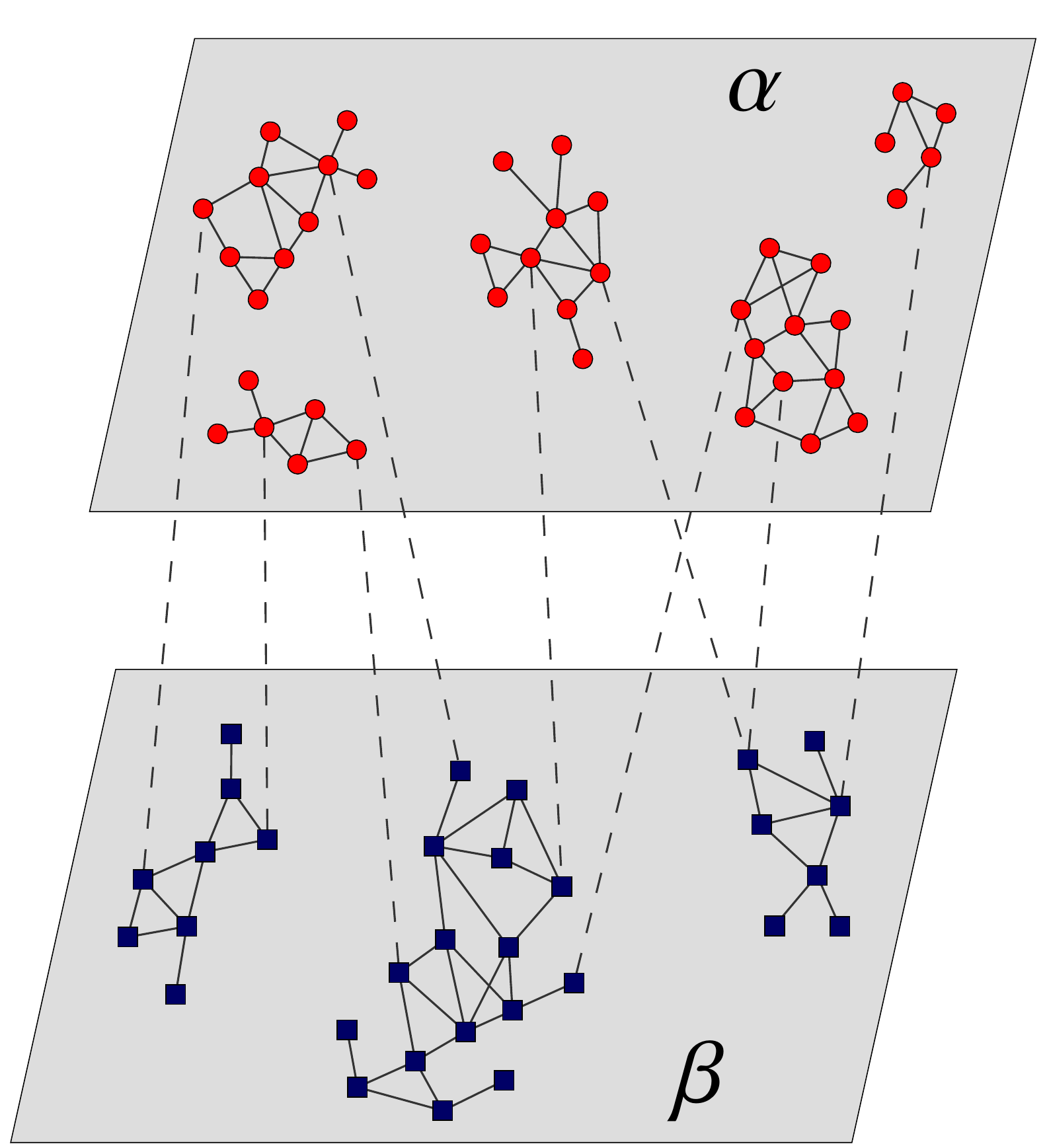}}
    \subfigure[]{\label{fig:excessDegree}
      \includegraphics[width=.25\textwidth]{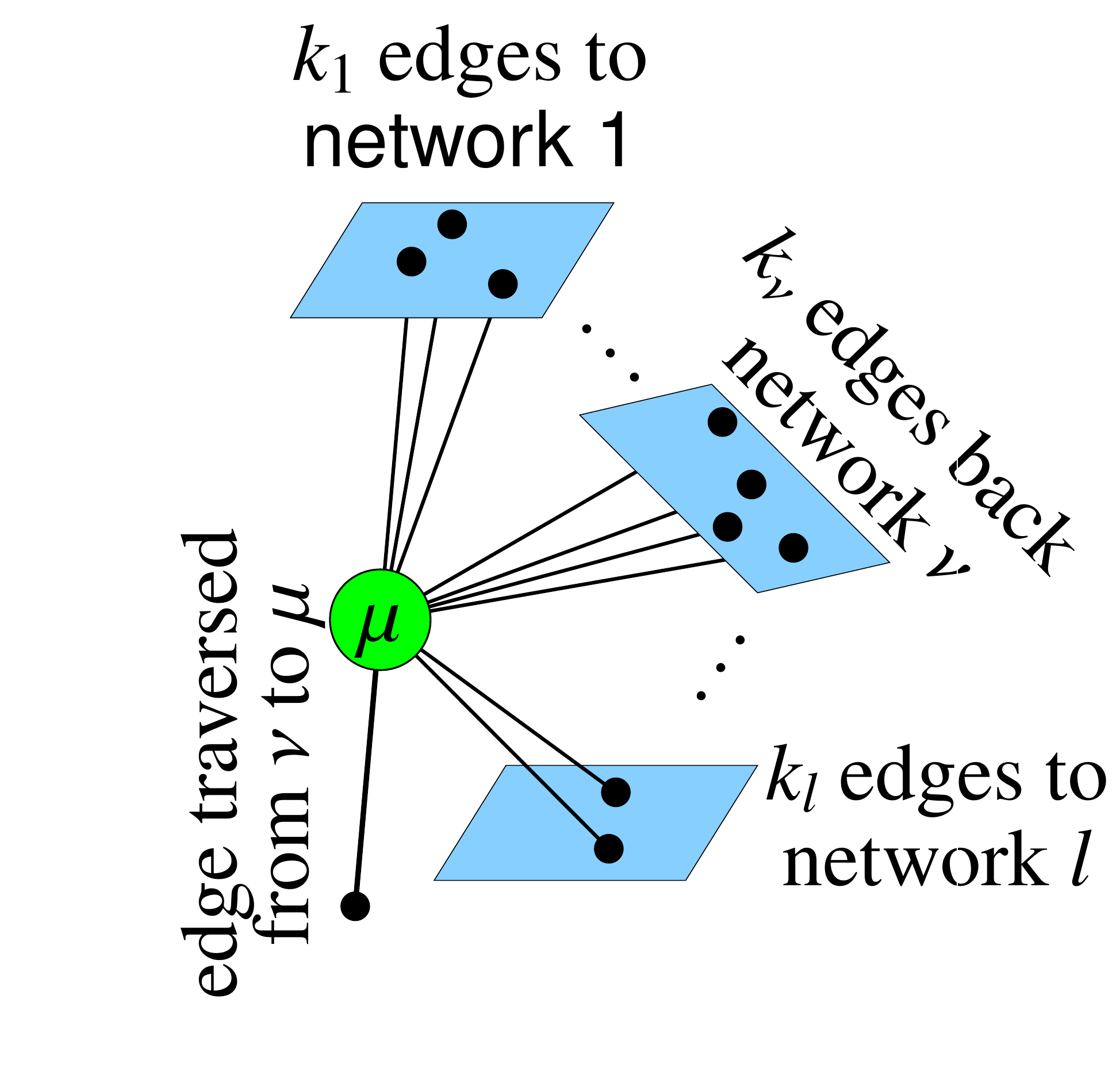}}
  \end{center}
  \vspace{-0.2in}
  \caption{a) Two networks $\alpha$ and $\beta$.  Nodes interact
    directly with other nodes in their immediate network, yet also
    with nodes in the second network. b) An illustration
    of the remaining edges incident to a node in a
    network \m~reached by following a random edge between
    networks \n~and \m.}
\label{fig:intro}
\end{figure}

Consider a system formed by two interacting networks,
\a~and \b, Fig~\ref{fig:2layers}.
Network \a~could be a human contact network for 
one geographic region and network \b~that for a separated region. 
When viewed as individual systems,
only small clusters of connected nodes exist, hence, a disease
spreading in either network should stay contained within clusters. 
In reality, 
a disease can hop from \a~to \b, for instance, by an infected person flying on a
airplane, spread in the \b~network and eventually hop back 
to the \a~network into new clusters, causing an epidemic outbreak. 
Next consider interacting networks that contain completely different types of nodes.  Network \a~can be a social network, such as an email communication network of software developers, while
network \b~can be a technological network, such as the network of calls between functions in software code. 
Here, bi-partite edges connect developers on $\alpha$ to code they author on $\beta$.


An important step towards modeling interacting networks was introduced with the 
layered network framework 
of~\cite{ThiranLayeredNetworks2006}. Yet, the networks in the distinct layers must be composed of the identical nodes (modeling essentially physical connectivity and logical connectivity or flow).  
Herein we consider systems of $l\geq2$ distinct interacting networks
and calculate explicitly how the connectivity
within and between networks determines the onset of large scale
connectivity in the system  and in each network
individually. Our mathematical formulation has some overlap with recent works 
calculating connectivity properties in a single network accounting for a diversity of node attributes~\cite{BolloJanRior07,PourboPRE09} or interactions between modules within a network~\cite{DoroMenPRE08,OstilliMendes}. 
Here we present our formalism and also applications to real-world systems  
of interacting networks coming from telecommunications and software. 

The onset of large-scale connectivity (i.e., the
percolation threshold) corresponding to the emergence of a giant
connected component in an isolated network has been studied
extensively, first for random networks with Poisson degree
distributions~\cite{ER1959} and later for random networks with
arbitrary degree distributions~\cite{MolloyReed1995}.
Similar results were then derived using 
generating functions~\cite{NewmanRandomGraphGF2001,CallawayRobustness2000},
the approach we employ herein.  
Generating functions, similar to the network configuration model~\cite{Bollobas1980, MolloyReed1995}, evaluate the ensemble of all possible
random networks consistent with a specified degree distribution,
$\{p_k\}$, and are most accurate in the sparse regime where networks
are approximately tree-like. Thus in the regime before the emergence of the giant
component, generating functions can be used to calculate the
distribution of component sizes.  In the supercritical regime they can
be used to calculate the distribution in sizes of components that are
not part of the giant component.

For our purposes, a system with $l\ge2$ interacting networks is
described by a set of degree distributions.  Each individual network
\m~is characterized by a multi-degree distribution,
$\{\pm_{k_1k_2\cdots  \kl}\}$, 
where $\pm_{k_1k_2\cdots \kl}$ is the
fraction of all nodes in network $\mu$ that have $k_1$
edges to nodes in network 1, $k_2$ edges to nodes in network 2,
etc.  The multi-degree distribution for each network may be written in
the form of a generating function:
\begin{equation}
\gm(x_1,\ldots, \xl) = \sum_{k_1,\ldots, \kl =
  0}^{\infty} \pm_{k_1\cdots \kl} x_1^{k_1}\cdots \xl^{\kl}.
\label{eq:mostGeneralGF}
\end{equation}
To simplify notation in what follows, we now define two $l$-tuple's,
$\vecx = (x_1, \ldots, \xl)$ and $\veco = (1,\ldots,1)$.

Our interest is in calculating the distribution of component sizes, 
where a component is a set of nodes connected to one another either
directly or indirectly by traversing a path along edges.  Clearly such
components can be composed of nodes distributed among the $l$
different networks, and our formulation allows us to calculate the
distribution of such system-wide components, yet also to refine the
focus and calculate the contribution coming from nodes contained in
only one of the $l$ networks.  

\begin{figure}[b]
\begin{center}
{\resizebox{3.4in}{!}{\includegraphics{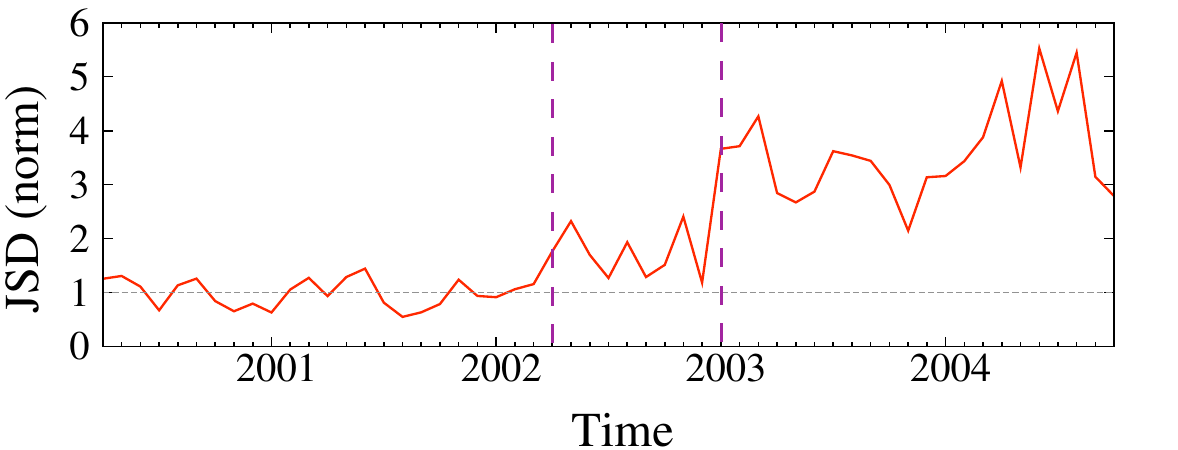}}}
\end{center}
\vspace{-0.1in}
\caption{Comparison over time of the distribution of the number of developers connected indirectly via co-editing code in the Apache project with the distribution expected in a random network with the same multi-degree distribution. Vertical lines mark the first generally available release in 2002, and a significant deviation from random in 2003, when the communication network shrinks and the project seems to become more efficiently organized.} 
\label{fig:apache}
\end{figure}

We begin by deriving the distribution of connectivity for
a node at the end of a randomly chosen edge.  
Consider selecting uniformly at random an edge falling between a node in
network \n~and a node in network \m\ (i.e., a \n-\m\ edge).  
The \m~node attached to the edge is $\kn$ times more likely to 
have \n-degree $\kn$ than degree 1.
We can also account for the remaining local connectivity, to nodes in other networks as shown in Fig.~\ref{fig:excessDegree}.  In single isolated networks remaining connectivity is called the \emph{excess degree} of a node~\cite{NewmanRandomGraphGF2001}.  
Let $\qmn_{k_1\cdots \kn \cdots \kl}$
denote the probability of following a randomly chosen \n-\m\ edge to a node with excess \n~degree as shown in Fig.~\ref{fig:excessDegree} (which has total \n-degree of $\kn+1$). Then $\qmn_{k_1\cdots \kn \cdots \kl}
\propto (\kn + 1) \pm_{k_1\cdots (\kn + 1)\cdots \kl}$, and the
generating function for the distribution, $\{\qmn_{\ko \cdots \kl}\}$
is,
\begin{eqnarray}
\label{eq:gmn}
\gmn &&\hspace{-0.12in}(\vecx) = \hspace{-0.1in}
\sum_{\ko, \ldots, \kl=0}^{\infty}\hspace{-0.0in}
\qmn_{\ko \cdots \kl} \xo^{\ko}\cdots\xl^{\kl}\\
\nonumber  &\hspace{-0.25in}=& \hspace{-0.2in}\sum_{\ko, \cdots
  \kl=0}^{\infty} \frac{(\kn + 1) \pm_{\ko \cdots (\kn + 1)\cdots
    \kl}} {\sum_{j_1, \ldots, j_l =0}^{\infty }(j_{\nu+1})\pm_{j_1
    \cdots (j_{\nu} + 1)\cdots  j_l}}\xo^{\ko} \cdots \xl^{\kl}  \\
\nonumber&\hspace{-0.25in}=& \hspace{-0.1in}\left(\sum_{j_1,\cdots,
  j_l=0}^{\infty} \hspace{-0.1in}j_{\nu} \pm_{j_1\cdots
  j_l}\right)^{-1} \hspace{-0.1in} 
\partxn\hspace{-0.05in}
\sum_{\ko,\ldots,\kl=0}^{\infty}  \hspace{-0.1in} \pm_{\ko\cdots\kl} 
\xo^{\ko}\cdots\xl^{\kl}\hspace{-0.05in}\\
\nonumber&\hspace{-0.25in}=& \frac{\gmdn(\vecx)}{\gmdn(\veco)} 
\end{eqnarray}
where $\gmdn(\vecx)$ denotes the first derivative of $\gm(\vecx)$ with
respect to $\xn$ and the denominator is a normalization constant so that 
$\gmn(\veco) = 1$.
Also note that $\gmdn(\veco)\equiv \dmn$
is the average \n-degree for a node in  network \m. 

The distribution of second nearest neighbors for that \m~node via the \n~layer is calculated by using Eq.~\ref{eq:gmn} as the argument to Eq.~\ref{eq:mostGeneralGF}, namely $\left.\gm(1,1,...,\gnm(\vecx)\right|_{x_{\lambda}=1,\lambda\ne\mu},...,1)$.  
Comparing this distribution calculated via generating functions to that found in real-world interacting networks can reveal interesting statistical features.
Returning to the software example, we have 
a network of email communication between developers,
a network of relations between code, and bipartite edges
connecting developers to the code they edit.  
We would expect that the real system 
does not resemble a random network, but instead reflects a structure conducive to project development. For instance, if two developers edit the same code we would like for them to directly communicate via email and thus be first neighbors.  
In a sparse random network these developers would typically be second neighbors, connected indirectly via the code they both edit. 

We analyze the evolution of the Apache 2.0 Open Source Software project from mid-2000 thru 2004, 
with data aggregated over three month windows. From this we extracted the multi-degree distribution 
of the system for each time-shot, which we then plug into 
our generating functions to calculate the expected distribution of second neighbors found by following first a developer-to-code edge then a code-to-developer edge.  We then compare
this distribution to the real distribution of such developer second nearest neighbors 
using the Jensen-Shannon divergence~\cite{JS}, a symmetric measure based on
Kullback-Leibler divergence.  
The results are shown in Fig.~\ref{fig:apache} with the
JS-score of the real networks normalized by the JS-scores from the
ensemble of random networks.  
Values greater or less than unity indicate networks more or less
random than average. We indicate two vertical bars where significant difference between the random and real networks occurs.  The first, in mid-2002 marks the first general availability release of Apache 2.0, the second, at the start of 2003, is a bug and security fix~\cite{apachewebpage}.  This latter point,
moreover, marks when a substantial purging of developers from the communication network occurs. 
In any three-month window we observe that only about 25 developers edit code, yet prior to 2003 the number of developers in the email network is significantly larger. Thus this time seems to indicate when the Apache project becomes more efficiently organized,
eliminating noise of spurious emails to inactive developers. 
%

\begin{figure}[tb]
{\resizebox{3.4in}{!}{\includegraphics{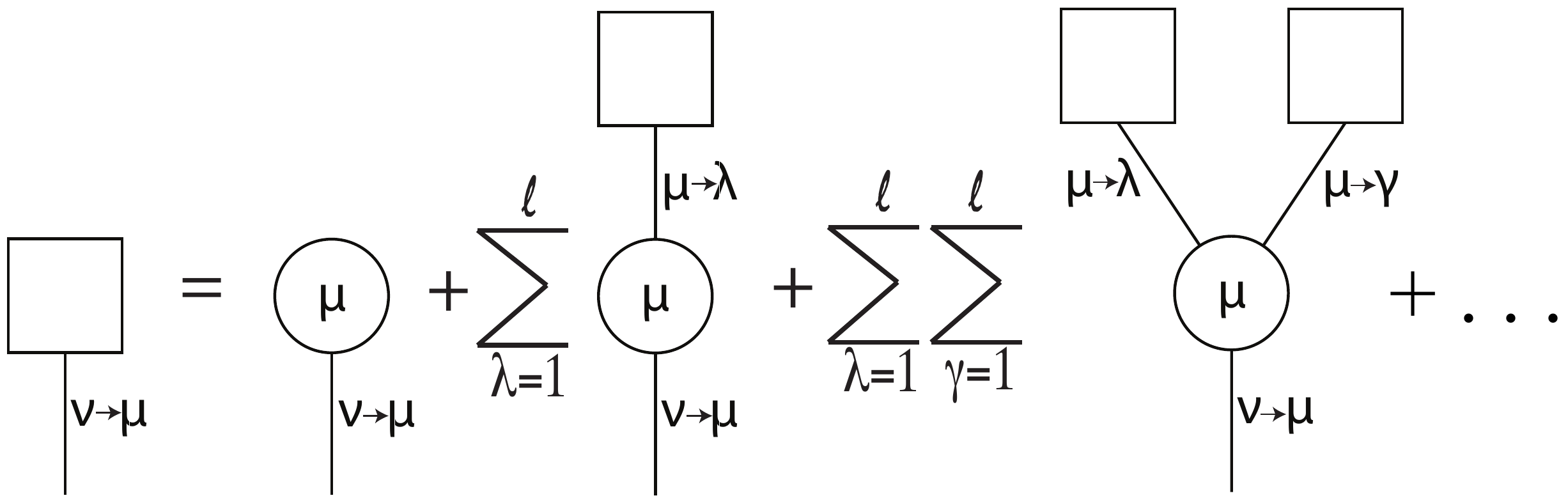}}}
\caption{A diagramatical representation of the topological constraints
  placed on the generating function $\hmn(\vecv)$ for the 
  distribution of sizes of components reachable by
  following a randomly chosen \n-\m\ edge. 
  The labels attached to each edge indicate type or
  \emph{flavor} of the edge and summation notation indicates that we are
  summing over all possible flavors.} 
\label{fig:componentTopo}
\end{figure}



We are now in position to consider component sizes. Assume we follow a
randomly chosen \n-\m\ edge to a \m~node (Fig.~\ref{fig:excessDegree}), 
and consider the distribution in sizes of
the component found by following the additional outgoing edges.
Let $\hmn(\vecv)$ denote the associated generating function.
Fig.~\ref{fig:componentTopo} illustrates all the types of connectivity possible for the \m-node, and summing over all these possibilities leads
to the self-consistency equation for $\hmn(\vecv)$:  
%
\begin{eqnarray}
\hmn&&\hspace{-0.25in}(\vecv)= \xm \qmn_{0\cdots0}\label{eq:hmn1}\\
\nonumber  &+&\xm\hspace{-0.15in} \sum_{\ko \ldots \kl =
  0}^{1}\hspace{-0.1in}  
\delta_{1,\sum_{\lambda=1}^lk_{\lambda}}\qmn_{\ko \cdots
  \kl}\prod_{\gamma = 1}^l H_{\gamma\mu}(\vecv)^{k_{\gamma}}\\
 &+&\xm\hspace{-0.15in}  \sum_{\ko,\ldots,\kl =
  0}^{2}\hspace{-0.1in}
\delta_{2,\sum_{\lambda=1}^lk_{\lambda}}\qmn_{\ko \cdots \kl}
\nonumber\prod_{\gamma = 1}^l H_{\gamma\mu}(\vecv)^{k_{\gamma}}+ \cdots
\end{eqnarray} 
$\delta_{ij}$ denotes the Kronecker delta, used here to 
account for all combinations of flavors of edges connected to the \m-node leading to specified excess degree $i$.  
Reordering the terms,  Eq.~\ref{eq:hmn1} becomes
\begin{equation}
\hmn(\vecv) =  \xm \sum_{\ko \ldots \kl =
  0}^{\infty}\qmn_{\ko \cdots \kl}
  \hom(\vecv)^{\ko}\cdots\hlm(\vecv)^{\kl}.
\end{equation}
We recognize the form of this equation from  Eq.~\ref{eq:gmn}, thus
\begin{equation}
\hmn(\vecv) = \xm\gmn[\hom(\vecv), \ldots,\hlm(\vecv)].
\label{eq:hmn2}
\end{equation}
We now consider starting from a randomly chosen \m-node, rather than a random \n-\m~edge. A topology such as one from Fig.~\ref{fig:componentTopo} exists a the
end of each edge incident to the \m-node.  The generating function
for the probability distribution of component sizes is,
\begin{equation}
\hm(\vecv) = \vm\gm[\hom(\vecv), \ldots, \hlm(\vecv)]\label{eq:hm}.
\end{equation}

While in theory it is possible to solve Eq.~\ref{eq:hmn2} for
$\hmn(\vecv)$ and use that solution in Eq.~\ref{eq:hm} to solve for
$\hm(\vecv)$, in practice, even for the case of a single isolated
network, as noted in~\cite{NewmanRandomGraphGF2001} the equations are
typically quite difficult 
to solve.  Yet, Eq.~\ref{eq:hm} allows
calculation of average component size.  A component may 
include multiple node flavors, but we can distinguish
between the average number of each type.  For example, the average
number of \n-nodes in the component of a randomly chosen \m-node is
\begin{eqnarray}
 \langle \sm\rangle_{\nu} &=&
\left.\partvn\hm(\vecv)\right|_{\vecv=\veco}\nonumber \\
\nonumber &=& \delta_{\mu\nu}\gm[\hom(\veco),\ldots,\hlm(\veco)]\\
\nonumber &&  +\sum_{\lambda = 1}^l
\gmdl[\hom(\veco),\ldots,\hlm(\veco)]\hlmdn(\veco)\\
&=& \delta_{\mu\nu} +\sum_{\lambda = 1}^l
\gmdl(\veco)\hlmdn(\veco)\label{eq:smn}
\end{eqnarray}
Intuitively Eq.~\ref{eq:smn} is reasonable because $\hgldn(\veco)$
represents the average number of $\nu$-nodes in the component found by
following a $\mu$-$\lambda$ edge towards a $\lambda$-node, and the
expected number of $\mu$-$\lambda$ edges incident to an initial \m-node
is $\gmdl(\veco)$ (recall, $\gmdl(\veco) =
\dml$).  The product of the two terms summed over all $\lambda$
networks produces the number of $\nu$-nodes in a component connected
to a randomly chosen $\mu$-node, $\langle \sm\rangle_{\nu}$.

The preceding results regarding components hold in the sub-critical
regime where no giant connected component exists.  Once a giant
component emerges, generating functions allow us to calculate
properties of components {\it not} belonging to it.  The giant
component will span multiple networks and calculating its size
requires accounting for the contribution from each network.
Let $\Sm$ be the fraction of \m-nodes belonging to the giant
component.  The probability that a randomly chosen \m-node is
\emph{not} part of the giant component must then satisfy the following
equation,
\begin{equation}
1 - \Sm =\hspace{-0.14in}
\sum_{\ko,\ldots,\kl=0}^{\infty}\hspace{-0.14in}\pm_{\ko,\ldots,\kl}\uom^{\ko}\cdots\ulm^{\kl}
= \gm(\uom,\ldots,\ulm),
\label{eq:S}
\end{equation}
where $\unm$ is the probability that an \m-\n\ edge is not part of the giant component.
In addition, for all $\mu,\nu\in l$, $\unm$ must satisfy,
\begin{equation}
\unm = \gnm(\uon,\ldots,\uln)\label{eq:unm},
\end{equation}
derived using the same self-consistency arguments that resulted in Eq.~\ref{eq:hmn2}.

Though all the equations above hold for a system of $l \ge 2$
interacting networks, we now give a concrete example for $l=2$, with
the networks indexed as \a~and \b.  Consider first the simplest of
systems, where the internal connectivity of \a~and 
\b~each has a distinct Poisson degree distribution, and the
inter-network connectivity is described by a third Poisson degree
distribution, for instance, $p^\alpha_{k_\alpha k_\beta} = \left[(\daa)^{\ka}e^{-\daa}/{\ka}!\right] \left[(\dab)^{\kb}e^{-\dab}/{\kb}!\right]$. (Recall $\dmn$ denotes the average \n-degree for a node in network \m.)
Then, from Eq.~\ref{eq:mostGeneralGF},
\begin{eqnarray}
\ga(\xa,\xb) &=& e^{\daa(\xa-1)}e^{\dab(\xb-1)}\label{eq:gaPoisson}\\
\gb(\xa,\xb) &=& e^{\dba(\xa-1)}e^{\dbb(\xb-1)}\label{eq:gbPoisson}.
\end{eqnarray}
Using Eq.~\ref{eq:smn}, the average number of \a-nodes in a component
reachable from a randomly chosen \a-node is,
\begin{equation}
\langle \ssa\rangle_{\alpha} = 1 + \frac{\daa + \dab\dba
  -\daa\dbb}{(1-\daa)(1-\dbb) - \dab\dba}.
\end{equation}
The average component size diverges for
$(~1~-~\daa~)~(~1~-~\dbb~)~=~\dab\dba$; the point at which the giant
component emerges. 
(Ref.~\cite{OstilliMendes} recently presented an alternate method for deriving similar percolation thresholds and connectivity properties, but in a single network with multiple interacting communities.)
%
Note, following Eq.~\ref{eq:smn}, we can show
$\langle \ssb\rangle_{\alpha}$, $\langle \ssa\rangle_{\beta}$,  and
$\langle \ssb\rangle_{\beta}$ also  all diverge at this point,
marking when 
a giant component emerges in each network and throughout the system.  Further simplifying, by assuming the two interacting networks have the
same degree distribution, $\daa = \dbb = \dintra$ and
$\dab=\dba=\dinter$, then the giant component emerges when, $\dinter +
\dintra = 1$, recovering the standard result for a single network
(which, by definition, has $\dinter = 0$) that emergence occurs for
$\dintra = 1$. 

\begin{figure}[tbh]
  \begin{center}
    \includegraphics[width=.5\textwidth]{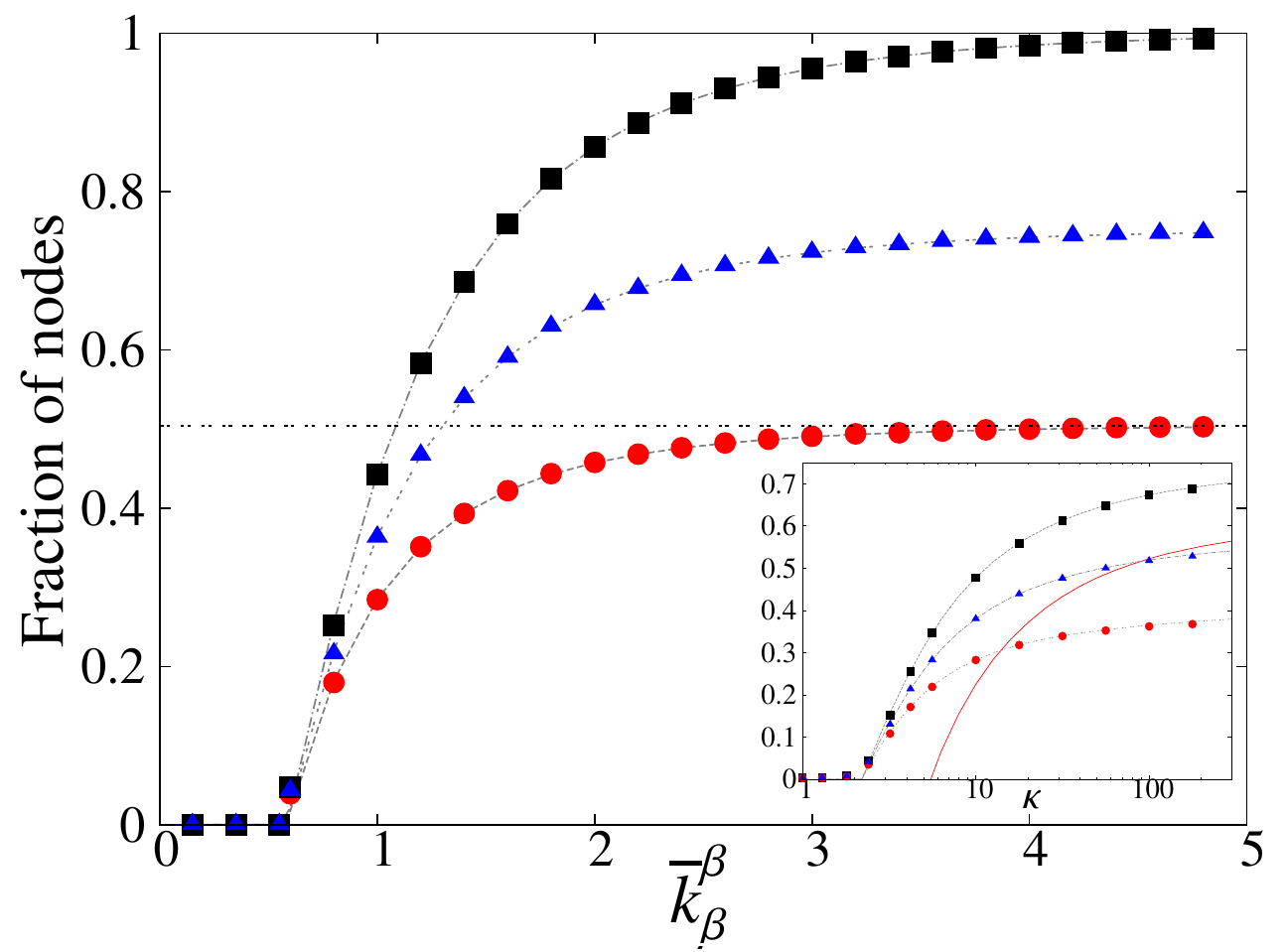}
  \end{center}
  \vspace{-0.1in}
  \caption{Numerical simulations of connectivity in a system of
    two interacting Poisson degree distributed networks, \a~and \b, with inter-network connectivity also   Poisson distributed, as connectivity on \b~increases. Each network has 100,00 nodes, with $\daa = 0.4$ and $\dab = \dba = 0.5$.  Shown are the fraction of \a~nodes, $S_\alpha$  (circles), \b~nodes,
$S_\beta$ (squares), and all nodes, $S$ (triangles) in the system-wide giant
    component, with the dashed curves giving the analytic results, Eqns.~(13) and (14). 
  The horizontal dashed line is the asymptotic value to which $S_\alpha$
  approaches.  (Inset)  Analogous results when \a~has Poisson distribution with $\daa = 0.5$, inter-network edges follow a  Poisson distribution with $\dab
  = \dba = 0.4$, but \b~has a power-law distribution with exponent  $\tau = 2.5$ and an
  exponential cutoff that we vary between $1 \leq \kappa \leq  300$.   
  The solid curve is the result for network \b~when viewed in isolation.}
\label{fig:Splot}
\end{figure}


Once the giant component emerges the $\unm$ which satisfy
Eq.~\ref{eq:unm} are 
$\uaa = \uab = 1 - \Sa$ and $\ubb = \uba
= 1-\Sb$, while $\Sa$ and $\Sb$, respectively, the number of \a-nodes and \b-nodes in the giant component of the system, satisfy
\begin{eqnarray}
\Sa = 1 - e^{-(\daa\Sa + \dab\Sb)}\label{eq:SaPoisson}\\
\Sb = 1 - e^{-(\dba\Sa + \dbb\Sb)}\label{eq:SbPoisson}.
\end{eqnarray}

To observe the change in connectivity of one network precipitated by
an increase in connectivity of a second network attached to the first,
we simulated  a system of two interacting networks and fixed $\daa$,
$\dab$, and $\dba$ while varying  $\dbb$ from 0 to 5
(Fig.~\ref{fig:Splot}).  As $\dbb$ 
increases the \b-network becomes a single connected component (the
traditional behavior for a single network) and $\Sb \rightarrow 1$.
However, the connectivity of \a~remains limited.  It can be shown that
as $\dbb$ increases $\Sa \rightarrow \frac{-1}{\daa}W\left[-\daa 
  e^{-\daa   -\dab}\right]+1$ (dashed horizontal line in Fig.~\ref{fig:Splot}), 
  where $W$ is the Lambert $W$ function, also known as the product log.

We next consider more complex degree distributions, where
\a~is still described by a Poisson distribution, but the internal
connectivity of \b~is described by a power-law distribution with an
exponential cutoff.  
While power-law degree distributions have
attracted considerable attention as a model for node degree
distributions in many types of networks~\cite{BA1999}, 
a power-law with an exponential cutoff
may be a better model for real-world networks~\cite{ClausetShaliziNewman2009}.
Here
$p^\beta_{k_\alpha k_\beta} = 
\left[(\dba)^{\ka}e^{-\dba}/{\ka}!\right]\left[(k_\beta)^{\tau}e^{-k_\beta/\kappa}/\textrm{Li}_{\tau}(e^{-1/\kappa})\right]$
where $\textrm{Li}_n(x)$ is the $n$th polylogarithm of $x$ and serves as
a normalizing factor for the distribution.  Thus, we can write our
basic generating function for network \b,
\begin{equation}
\gb(\xa,\xb) = e^{\dba(\xa-1)} \frac{\textrm{Li}_{\tau}(x_\beta e^{-1/\kappa})}
{\textrm{Li}_{\tau}(e^{-1/\kappa})}.
\label{eq:gbPL}
\end{equation}
The generating function for \a~is still given by
Eq.~\ref{eq:gaPoisson}.  We simulate the impact on the connectivity of
the \a-network as the exponential cutoff and hence the average degree
of network \b~increases, inset of Fig.~\ref{fig:Splot}.  Again the dashed curves are the analytic results obtained by solving Eqns.~\ref{eq:S} and \ref{eq:unm}. The solid red line is the behavior for the \b~network considered in isolation, showing that even the percolation threshold for \b~is lowered through connectivity with network \a.



\begin{figure}[t]
  \begin{center}
      \includegraphics[width=.45\textwidth]{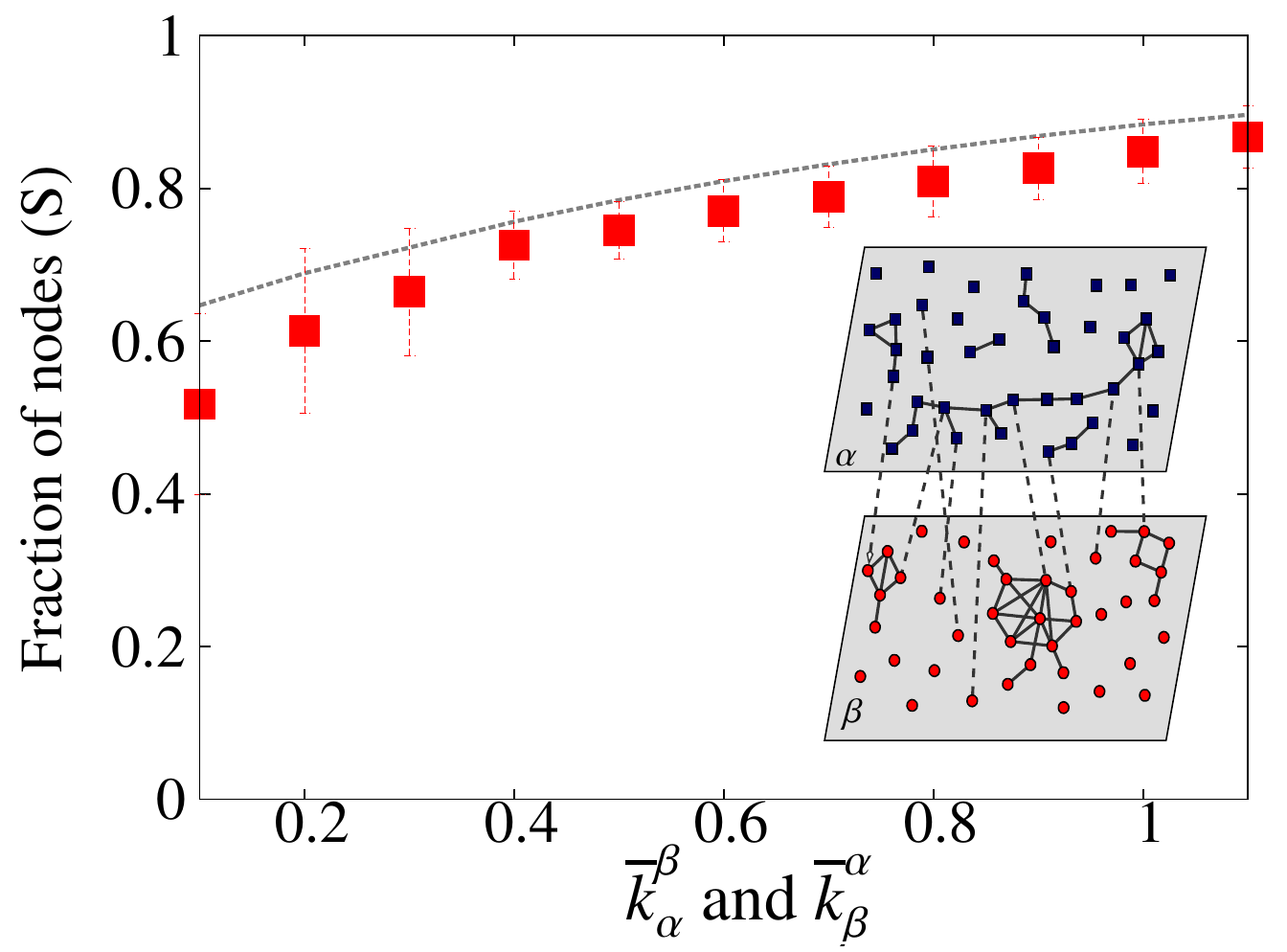}]
  \end{center}
  \vspace{-0.2in}
  \caption{Inset are two sample networks of Bluetooth connectivity. The main figure shows the increase in participation in the giant component as connectivity between \a~and \b~increases, starting from $\dab=\dba=0.1$. Points are obtained by taking the empirical data and simulating inter-network edges with the appropriate $\dab$ and $\dba$, averaged over 100 realizations. The solid line is from analytic calculations.}
\label{fig:iMotes}
\end{figure}

Finally we consider an application of connectivity to communications networks, building on the increasing interest in using Bluetooth connectivity between  individuals to transmit data~\cite{Ioannidis2008mobilePhone}.  For instance, rather than downloading a webpage (such as the CNN homepage) by connecting to the Internet, a copy could be obtained from a close-by individual already in possession of this data. We construct prototypical networks of local Bluetooth connectivity between individuals from raw data of Bluetooth sightings by 41 attendees at the 25th IEEEE International Conference on Computer Communications (INFOCOM)~\cite{imote}.  We initially partition the raw data into discrete 20 minute windows and consider that a communication edge exists between any two devices so long as they are within contact for at least 120 seconds.  Each network has approximately a Poisson degree distribution of connectivity. We choose two arbitrary 20 minute snapshots as proxies for two distinct networks, \a~and \b, representing, for instance, two separate rooms at the conference.   
We calculate how adding long-range connections between \a~and \b~(for instance via text messages or email) enhances overall connectivity in the system. 
%
In other words, we calculate how many long-range connections would be needed between two isolated local Bluetooth networks to create the desired large scale connectivity, 
potentially allowing many users to share information. Figure~\ref{fig:iMotes} shows the size of the giant component obtained via numerical simulations using the real data (points) and the analytic calculations obtained via generating functions (dashed line).  The analytic calculations slightly overestimate connectivity, yet there is remarkable agreement with empirical data  
even  though the actually networks are quite small. 


In summary, we have introduced a formalism for calculating
connectivity properties in a system of $l$ interacting networks.  We
demonstrate the extreme lowering of the percolation threshold possible
once interactions with other networks are taken into account.  This
framework for calculating connectivity and statistics of interacting networks should be broadly applicable, and we show potential applications to software and communications systems. 


{\bf Acknowledgements} We thank Christian Bird for providing data on the Apache project and for useful conversations.

\bibliographystyle{apsrev}
\bibliography{journals,multiNets}

\end{document}